\documentclass[5p,times]{elsarticle}
\usepackage{amsmath,amssymb}
\usepackage{graphicx}
\usepackage{booktabs}
\usepackage{float}
\usepackage{microtype}
\usepackage{xurl}
\usepackage[hidelinks]{hyperref}
\hypersetup{pdftitle={Solvent Inertia Changes Product Probability by Relocating the Phase-Space Reactivity Boundary},pdfauthor={Stephen Wiggins}}
\journal{Chemical Physics Letters}

\begin{document}
\begin{frontmatter}
\title{Solvent Inertia Changes Product Probability\\by Relocating the Phase-Space Reactivity Boundary}
\author[himis,bristol]{Stephen Wiggins\corref{cor1}}
\ead{S.Wiggins@bristol.ac.uk}
\affiliation[himis]{organization={Hetao Institute of Mathematics and Interdisciplinary Sciences},city={Shenzhen},country={China}}
\affiliation[bristol]{organization={School of Mathematics, University of Bristol},addressline={University Walk},city={Bristol},postcode={BS8 1TW},country={United Kingdom}}
\cortext[cor1]{Corresponding author.}
\begin{abstract}
Changing solvent inertia changes the kinetic part of a molecular Hamiltonian and hence its flow in phase space, without requiring a change in the potential energy surface (PES). We study how this affects product probability in a solute--solvent model with two degrees of freedom. Initial conditions are chosen at the same total energy for each solvent mass. For this family, the deterministic outcome changes from reaching products to returning to reactants at a threshold mass. The initial condition at that mass belongs to the stable manifold of the transition-state periodic orbit. Weak Gaussian forcing of the solvent momentum replaces the abrupt change in outcome by a continuous change in product probability over a mass interval proportional to the square root of the noise strength. The deterministic threshold and a linear response calculation predict the position and width of this probability change without fitting the stochastic simulations. Monte Carlo results at three noise strengths agree with the prediction within sampling uncertainty. The calculation connects solvent inertia to product probability through the geometry of the full Hamiltonian flow, rather than through a change in the PES.
\end{abstract}
\end{frontmatter}

\section{Introduction}
\label{sec:introduction}

Changing a mass in a molecular Hamiltonian changes its kinetic energy function and therefore its flow in phase space, even when the potential energy surface (PES) is unchanged. The question is not whether solvent inertia can affect reaction dynamics without altering the PES, but how the resulting change in reactive transport determines a product probability.

Garcia-Meseguer and Carpenter introduced a model of a reacting solute coupled to a collective solvent coordinate to isolate an inertial contribution to solvent effects \cite{GarciaCarpenter2019}. Garcia-Meseguer, Carpenter, and Wiggins subsequently studied its phase space dynamics \cite{GarciaCarpenterWiggins2019}. The effective solvent mass enters only through the kinetic part of the Hamiltonian. Varying that mass relocates the transition-state periodic orbit and its associated dividing surface while leaving the PES unchanged. We now ask how this deterministic geometry determines the probability of reaching products when weak fluctuations act through the solvent coordinate.

We formulate this question using transition-state theory in phase space. For Hamiltonian systems with two degrees of freedom, Pechukas, Pollak, and their collaborators developed the construction based on unstable periodic orbits and their dividing surfaces \cite{PechukasMcLafferty1973,PollakPechukas1978,PechukasPollak1979}. Normally hyperbolic invariant manifolds provide the corresponding structures in systems with more degrees of freedom \cite{Uzer2002,Waalkens2008}. In the present model, two objects associated with the transition-state periodic orbit have different roles. Its dividing surface is crossed by reactive trajectories; its stable manifold locally separates initial conditions with different deterministic outcomes. We use the latter as the deterministic reactivity boundary.

For each solvent mass we choose an initial condition with the same positions, solute momentum, and total energy. The solvent momentum is adjusted to satisfy the energy constraint. For this family there is a threshold mass separating trajectories that reach the product threshold first from those that reach the reactant threshold first. Weak random forcing makes both outcomes possible near this mass. Our calculation predicts where the product probability changes and the mass interval over which it changes, using the deterministic threshold and the response to the specified solvent forcing.

Kramers and Grote--Hynes theory already show that solvent dynamics can affect reaction rates through transmission and recrossing \cite{Kramers1940,GroteHynes1980}. We do not propose mass dependence at fixed PES as a new principle. We study a specified family of initial conditions rather than an equilibrium rate, and ask whether the reactivity boundary provides a quantitative description of its product probability. The distinction is not between potential energy and dynamics, but between information contained in the PES alone and information contained in the full Hamiltonian flow.

\section{Solvent inertia and the deterministic reactivity boundary}
\label{sec:model}

Let $r_1$ be the solute coordinate, $r_2$ the collective solvent coordinate, and $p_1,p_2$ their conjugate momenta. The Hamiltonian is \cite{GarciaCarpenterWiggins2019}
\begin{equation}
H_{\mu_2}=\frac{p_1^2}{2\mu_1}+\frac{p_2^2}{2\mu_2}+V(r_1,r_2),
\label{eq:H}
\end{equation}
where the solute mass is $\mu_1=1$, the effective solvent mass is $\mu_2$, and
\begin{equation}
V(r_1,r_2)=\sum_{j=1}^{5}c_jr_1^{j-1}+c_6(c_7-r_2)^2+\frac{c_8}{(r_2-r_1)^{12}}.
\label{eq:V}
\end{equation}
The coefficients are
\begin{align*}
c_1&=321.904484, & c_2&=-995.713452,\\
c_3&=1118.689573, & c_4&=-537.856726,\\
c_5&=92.976121, & c_6&=1,\\
c_7&=1, & c_8&=0.01.
\end{align*}
The solvent mass is the only Hamiltonian parameter varied. All numerical values are in the units of this model.

We use the energy of the original phase space study \cite{GarciaCarpenterWiggins2019},
\begin{equation}
E=3.691966889.
\label{eq:E}
\end{equation}
The PES saddle has coordinates and potential energy
\begin{equation}
\begin{split}
(r_1^\ddagger,r_2^\ddagger)&=(1.36560508,2.16176887),\\
V^\ddagger&=3.47291422,
\end{split}
\label{eq:saddle}
\end{equation}
so $E-V^\ddagger=0.21905267$. The solute coordinates of the reactant and product minima are, respectively,
\begin{equation}
r_{1,R}=0.98778676,\qquad r_{1,P}=1.98516984.
\label{eq:thresholds}
\end{equation}
We use these as stopping thresholds. A trajectory is assigned the product outcome if $r_1$ reaches $r_{1,P}$ before $r_{1,R}$, and the reactant outcome if it reaches $r_{1,R}$ first. This classification concerns which threshold is reached first, not the subsequent residence of the trajectory in a well.

The earlier study sampled dividing surfaces \cite{GarciaCarpenterWiggins2019}. Here we choose one initial condition for each solvent mass in order to follow a particular change in outcome. The initial positions are
\begin{equation}
(r_1(0),r_2(0))=(r_1^\ddagger,r_2^\ddagger).
\label{eq:q0}
\end{equation}
At this configuration, the largest positive solute momentum compatible with energy $E$ occurs when $p_2=0$. We set
\begin{equation}
p_1(0)=0.75\sqrt{2(E-V^\ddagger)}
\label{eq:p1}
\end{equation}
and take the negative solvent momentum satisfying the energy constraint,
\begin{equation}
p_2(0;\mu_2)=-\sqrt{2\mu_2\left[E-V^\ddagger-\frac{p_1(0)^2}{2}\right]}.
\label{eq:p2}
\end{equation}
The factor $0.75$ specifies the family; it has no special physical significance. It assigns $0.75^2$ of the initial kinetic energy to the solute term and the remainder to the solvent term. Thus both the initial total energy and its division between the kinetic terms are the same for every mass. The solvent momentum and velocity are not: their magnitudes vary as $\sqrt{\mu_2}$ and $1/\sqrt{\mu_2}$, respectively.

Write $x_0(\mu_2)$ for the initial point defined by Eqs.~\eqref{eq:q0}--\eqref{eq:p2}. By construction,
\begin{equation}
H_{\mu_2}\bigl(x_0(\mu_2)\bigr)=E.
\label{eq:initial-energy}
\end{equation}
Each deterministic trajectory conserves this energy. We are comparing different Hamiltonians on their respective energy surfaces with the same energy value, not holding a phase space point fixed while changing its mass.

Starting with nearby masses giving opposite outcomes, bisection in $\mu_2$ gives
\begin{equation}
\mu_c\simeq9.975918.
\label{eq:muc}
\end{equation}
This is the threshold studied below. Other initial conditions can give a different threshold, and the calculation does not establish uniqueness over all solvent masses.

For a fixed mass, let $N_{\mu_2,E}$ denote the transition-state periodic orbit at energy $E$, and $W^s(N_{\mu_2,E})$ its stable manifold. At the threshold \cite{Wiggins2026},
\begin{equation}
x_0(\mu_c)\in W^s(N_{\mu_c,E}).
\label{eq:incidence}
\end{equation}
The deterministic trajectory from this point remains in the stable manifold. The change in outcome is obtained by comparing $x_0(\mu_2)$ with $W^s(N_{\mu_2,E})$ at different masses. It is not a crossing of an invariant manifold by a deterministic trajectory at fixed mass.

At fixed energy, the stable manifold is two-dimensional within the three-dimensional energy surface. The forcing introduced below changes energy, so we need the corresponding boundary across nearby energy surfaces. For a small energy interval $I$ containing $E$, consider
\begin{equation}
\mathcal N_{\mu_2,I}=\bigcup_{E'\in I}N_{\mu_2,E'}.
\label{eq:cylinder}
\end{equation}
Where the family is smooth and uniformly normally hyperbolic, this union is a two-dimensional cylinder with a three-dimensional stable manifold in the four-dimensional phase space \cite{Wiggins2026}. Intersecting that stable manifold with $H_{\mu_2}^{-1}(E)$ recovers the fixed-energy boundary locally. We use this extension as the deterministic reactivity boundary for the forced problem.

\section{Product probability under weak solvent forcing}
\label{sec:prediction}

We add a random force to the solvent momentum:
\begin{equation}
\begin{split}
dr_1&=p_1\,dt,\qquad dr_2=\frac{p_2}{\mu_2}\,dt,\\
dp_1&=-\partial_{r_1}V\,dt,\\
dp_2&=-\partial_{r_2}V\,dt+\sqrt{\varepsilon}\,dW_t,
\end{split}
\label{eq:sde}
\end{equation}
where $W_t$ is a standard Wiener process and $\varepsilon$ sets the noise strength. At a given mass, every realization starts at the same point $x_0(\mu_2)$; the random forces differ. We write $P_{\rm prod}(\mu_2,\varepsilon)$ for the probability of reaching the product threshold before the reactant threshold in Eq.~\eqref{eq:thresholds}.

We first consider the deterministic problem near $\mu_c$. The initial point and its reactivity boundary depend smoothly on mass. At the threshold considered here their signed separation has a nonzero derivative with respect to mass, so its first-order change is proportional to $\mu_2-\mu_c$. Increasing the mass favors return to reactants for this family.

The unforced solution starting at $x_0(\mu_c)$ provides the reference for the noise calculation. Linearizing the response to the Gaussian force gives a Gaussian displacement transverse to the reactivity boundary. Its standard deviation is proportional to $\sqrt{\varepsilon}$. The deterministic separation and the stochastic displacement are measured with respect to the same boundary. Their relative size gives a probability law depending on $(\mu_2-\mu_c)/\sqrt{\varepsilon}$.

The sensitivity calculation along the threshold solution gives \cite{Wiggins2026}
\begin{equation}
\kappa=0.138291.
\label{eq:kappa}
\end{equation}
This is the quantity denoted $\sqrt{2C}$ in Ref.~\cite{Wiggins2026}; the calculation and its implementation are described in Supplementary Material Sec.~\ref{sec:si-coefficient} and in the Zenodo archive \cite{Zenodo2026}. It combines the change in separation caused by varying the mass with the response to the prescribed solvent force. The prediction is
\begin{equation}
P_{\rm prod}(\mu_2,\varepsilon)\simeq
\Phi\!\left[-\kappa\frac{\mu_2-\mu_c}{\sqrt{\varepsilon}}\right],
\label{eq:prob}
\end{equation}
where $\Phi$ is the cumulative distribution function of a Gaussian with mean zero and variance one. The minus sign reflects the change toward the reactant outcome as mass increases.

The center of this probability change is fixed by the deterministic threshold, and its width by $\sqrt{\varepsilon}/\kappa$. Neither is fitted to the simulated probabilities. The approximation describes motion near the reactivity boundary. Noise can still alter the outcome before a stopping threshold is reached; agreement with the probability of the complete trajectory does not follow from the local calculation alone. We therefore test Eq.~\eqref{eq:prob} by following the full stochastic trajectories to the two stopping thresholds.

\section{Numerical verification}
\label{sec:results}

We compare Eq.~\eqref{eq:prob} with simulations at the three noise strengths
\[
\varepsilon=10^{-5},\qquad2.5\times10^{-6},\qquad10^{-6}.
\]
This tests both the dependence on solvent mass and the predicted square-root dependence of the width on noise strength. The equations were integrated by symmetric drift--diffusion splitting, with fourth-order Runge--Kutta for the Hamiltonian substep and time step $\Delta t=10^{-3}$. Each probability was estimated from 30,000 independent trajectories followed until one of the two thresholds was reached. The algorithm, sampling grid, and time-step tests are given in Supplementary Material Secs.~\ref{sec:si-methods}, \ref{sec:si-grid}, and~\ref{sec:si-dt}.

\begin{figure*}[t]
\centering
\includegraphics[width=0.96\textwidth]{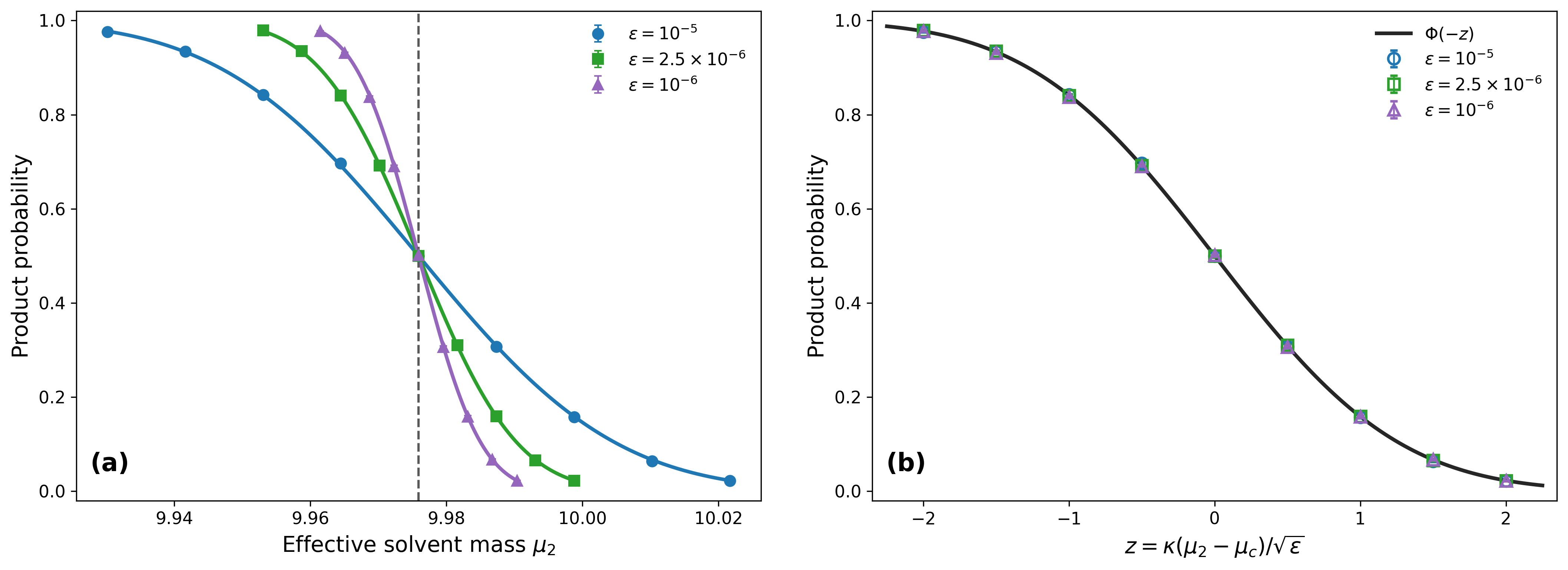}
\caption{Product probability and its dependence on solvent inertia. (a) Simulated probabilities for three noise strengths, with the prediction in Eq.~\eqref{eq:prob}. The dashed line marks the deterministic threshold mass. (b) The same probabilities plotted against $z$ in Eq.~\eqref{eq:z}. The curve is $\Phi(-z)$, with no fitted center or scale. Colors and marker shapes identify the same noise strengths in both panels. Error bars show binomial standard errors and are smaller than the symbols where they are not visible.}
\label{fig:collapse}
\end{figure*}

Figure~\ref{fig:collapse}(a) shows the product probability against solvent mass. As the noise decreases, the change from predominantly product to predominantly reactant outcomes occurs over a narrower mass interval, centered near the deterministic threshold. At $\varepsilon=10^{-5}$, increasing the mass from $9.9302$ to $10.0217$ changes the product probability from $0.976$ to $0.022$. All initial conditions have the same total energy, and the PES is unchanged. The difference lies in the resulting Hamiltonian flow and its response to solvent forcing.

To compare the three noise strengths on the scale predicted by Eq.~\eqref{eq:prob}, we use
\begin{equation}
z=\kappa\frac{\mu_2-\mu_c}{\sqrt{\varepsilon}}.
\label{eq:z}
\end{equation}
The prediction is that all three data sets follow the specified curve $\Phi(-z)$, without adjusting their centers or widths. This tests more than whether the data can be made to coincide: the horizontal scale is fixed by the independently calculated $\kappa$. Figure~\ref{fig:collapse}(b) shows agreement with that curve. Across the 27 probabilities, the root-mean-square difference is $1.73\times10^{-3}$, consistent with sampling uncertainty alone ($\chi^2=26.2$ for 27 probabilities with no fitted parameters, $p=0.51$; Supplementary Material Sec.~\ref{sec:si-grid}).

A second test holds the mass at $\mu_2-\mu_c=0.05$, on the deterministic reactant side, and reduces the noise strength. At the smallest noise used, the measured product probability is $9.3\times10^{-4}$, compared with $9.9\times10^{-4}$ from Eq.~\eqref{eq:prob}. The agreement thus extends to approximately one product event per thousand trajectories. This test and its figure are in Supplementary Material Sec.~\ref{sec:si-tail}.

\section{Discussion and conclusions}
\label{sec:discussion}

The earlier deterministic study established that solvent inertia relocates the periodic-orbit dividing surface in this model \cite{GarciaCarpenterWiggins2019}. Here the stable manifold associated with that orbit supplies a prediction for product probability under weak forcing. It identifies the mass at which the chosen initial conditions change deterministic outcome, and the response near that threshold determines the width of the probability change. The simulations test the connection between this geometry and the outcome of complete trajectories.

Holding the initial energy fixed does not hold the dynamics fixed. The relation between momentum and velocity changes with solvent mass, and the coupled solute and solvent motions evolve under different Hamiltonian flows. The potential energy is part of each of those flows; it has not been removed from the mechanism. What the phase space calculation supplies is the relation between the chosen initial conditions, the reactivity boundary, and the response to solvent forcing. A stationary-point potential energy difference does not supply that relation.

This is a probability conditioned on specified initial conditions, not an equilibrium reaction rate or a final product yield. Kramers and Grote--Hynes theory already account for dynamical contributions to rates \cite{Kramers1940,GroteHynes1980}. The present calculation concerns the geometry of a different observable. Its conclusion is limited to the neighborhood of the threshold and the forcing regime tested here, rather than to all solvent masses or all initial conditions.

The additive force is not a complete equilibrium solvent bath. Its mean energy input is small on the observed transit times; the estimate is given in Supplementary Material Sec.~\ref{sec:si-energy}. There is also a connection to a Langevin bath in the weak coupling limit at fixed temperature. Fluctuation--dissipation balance then makes friction proportional to the noise strength. Over bounded transit times its contribution is smaller than the displacement proportional to the square root of the noise strength, so the Hamiltonian threshold and $\kappa$ give the leading weak coupling prediction. Finite friction requires a dissipative treatment, such as the stochastic transition-state constructions of Bartsch, Hernandez, and Uzer \cite{BartschPRL2005,BartschJCP2005}.

The model does not establish the importance of this mechanism in a particular liquid. It does show how solvent inertia enters a calculation of product probability through the full Hamiltonian flow. The phase space description is useful here because it predicts an observable consequence of the solvent response, rather than merely identifying a geometrical structure.

\section*{CRediT authorship contribution statement}
Stephen Wiggins: Conceptualization, Methodology, Formal analysis, Investigation, Validation, Visualization, Writing -- original draft, Writing -- review and editing.

\section*{Data and code availability}
The numerical data, simulation and verification scripts, and code for the deterministic sensitivity calculation are archived on Zenodo, version 1.0.0, at \url{https://doi.org/10.5281/zenodo.22852306} \cite{Zenodo2026}. The Supplementary Material gives the numerical methods and statistical checks. The archive contains aggregate counts and probabilities, together with the seeds and code needed to regenerate the trajectories; individual trajectory time series are not stored.

\section*{Acknowledgments}
The author thanks R. Garcia-Meseguer and B. K. Carpenter for their collaboration on the earlier phase space study of the solvent-inertia model that motivated this work.

\section*{Declaration of competing interest}
The author declares that he has no known competing financial interests or personal relationships that could have appeared to influence the work reported in this paper.

\section*{Declaration of generative AI and AI-assisted technologies in the manuscript preparation process}
During the preparation of this work the author used Claude (Anthropic) and ChatGPT (OpenAI) to assist with manuscript organization, language, and revision, and with checking statistical and numerical claims. The author reviewed and edited the output, verified the reported quantities against the underlying data and scripts, and takes responsibility for the content of the publication. Their supporting role in the numerical work is described in Supplementary Material Sec.~\ref{sec:si-repro}.

\clearpage
\onecolumn
\newgeometry{margin=1in}
\fontsize{11}{13.2}\selectfont
\setlength{\parindent}{1em}
\setlength{\parskip}{0pt}
\setcounter{section}{0}
\setcounter{subsection}{0}
\setcounter{equation}{0}
\setcounter{figure}{0}
\setcounter{table}{0}
\renewcommand{\thesection}{S\arabic{section}}
\renewcommand{\theequation}{S\arabic{equation}}
\renewcommand{\thefigure}{S\arabic{figure}}
\renewcommand{\thetable}{S\arabic{table}}
\renewcommand{\theHsection}{supp.\arabic{section}}
\renewcommand{\theHequation}{supp.\arabic{equation}}
\renewcommand{\theHfigure}{supp.\arabic{figure}}
\renewcommand{\theHtable}{supp.\arabic{table}}
\begin{center}
{\Large\bfseries Supplementary Material}\par\medskip
{\large Solvent Inertia Changes Product Probability by Relocating the Phase-Space Reactivity Boundary}\par\medskip
Stephen Wiggins
\end{center}
\pdfbookmark[0]{Supplementary Material}{supplement}
The data and code supporting these calculations are archived on Zenodo, version 1.0.0 \cite{Zenodo2026}. This supplement describes the numerical experiments, the uncertainty estimates, and the checks of the reported results.

\section{Numerical integration and stopping rules}
\label{sec:si-methods}

We use the Hamiltonian and initial conditions in Sec.~\ref{sec:model} of the Letter. The displayed parameter values are
\[
E=3.691966889,\qquad
(r_1^\ddagger,r_2^\ddagger)=(1.36560508,2.16176887),\qquad
V^\ddagger=3.47291422,
\]
with stopping thresholds $r_{1,P}=1.98516984$ and $r_{1,R}=0.98778676$. The momenta are
\[
p_1(0)=0.75\sqrt{2(E-V^\ddagger)},\qquad
p_2(0;\mu_2)=-\sqrt{2\mu_2\left[E-V^\ddagger-\frac{p_1(0)^2}{2}\right]}.
\]
The code solves the potential-gradient equations for the minima and saddle and uses the resulting coordinates at greater precision than displayed here. Substituting the rounded printed coordinates is therefore not an exact replay of the archived calculation. Each initial condition has energy $E$; the initial shares of kinetic energy are $0.75^2$ in the solute term and $1-0.75^2$ in the solvent term.

The stochastic equations are Eq.~\eqref{eq:sde} of the Letter. A time step consists of an independent Gaussian increment to $p_2$ with variance $\varepsilon\Delta t/2$, a full deterministic step of length $\Delta t$ using fourth-order Runge--Kutta, and a second independent Gaussian increment with the same variance. This is a symmetric drift--diffusion splitting. The fourth-order description applies to the deterministic substep, not to the stochastic method as a whole. Gaussian increments are generated by the archived SplitMix64/Box--Muller implementation, using a separate stream for each trajectory.

After each full step, the code first checks whether $r_1\geq r_{1,P}$ and then whether $r_1\leq r_{1,R}$. It stops at the first satisfied condition. These discrete tests approximate the continuous-time first-arrival event; the stopping time is not interpolated between steps. The calculations for the main probability curves use $\Delta t=10^{-3}$ and a maximum integration time of $7$. Section~\ref{sec:si-dt} gives the time-step comparison.

A numerical guard also stops an integration if the separation at the end of a step satisfies $r_2-r_1<0.15$. The original implementation stored such an exit with the reactant code. An instrumented replay of all 2.01 million trajectories, using the archived seeds and unchanged numerical substeps, reproduced every archived product count and recorded no guard activations, non-finite-state exits, or unresolved trajectories. The smallest recorded separation was $0.59003$. This is a check at step endpoints, not a bound on intermediate Runge--Kutta stages or continuous-time trajectories. The archive includes the replay records and a runner that records numerical failures separately from arrivals at the physical thresholds.

\section{Probability law and sensitivity coefficient}
\label{sec:si-coefficient}

The probability prediction is
\[
P_{\rm prod}(\mu_2,\varepsilon)\simeq\Phi(-z),\qquad
z=\kappa\frac{\mu_2-\mu_c}{\sqrt{\varepsilon}}.
\]
The constants stored in the calculation are
\[
\mu_c=9.975917605480646,\qquad
\kappa=0.13829137141820574.
\]
These digits identify the numerical inputs; they are not an error bound for the exact threshold or coefficient.

The deterministic sensitivity calculation is described in Ref.~\cite{Wiggins2026}, where $\kappa$ is written as $\sqrt{2C}$ with $C=0.009562251704364068$. The notation $C$ is used there for the quadratic action coefficient. The Letter uses $\kappa$ because it enters the probability law directly. No stochastic probability is used to determine either $\mu_c$ or $\kappa$.

The archived script \path{src/recompute_critical_coefficient.py} calculates these quantities from the model potential. It uses the mass bracket $[9.975,9.976]$, deterministic integration and bisection for the threshold, and shooting and variational/adjoint equations for the periodic orbit and its response. Four terminal times, $1.50$, $1.55$, $1.60$, and $1.65$, are used to check the coefficient. The executable wrapper is \path{scripts/recompute_coefficient.py}. This is the calculation required for the Letter, not a reproducer for every result in Ref.~\cite{Wiggins2026}.

\section{Sampling grid and comparison with the predicted probabilities}
\label{sec:si-grid}

At each of the three noise strengths, we use nine values $z=-2,-1.5,\ldots,2$ and set
\[
\mu_2=\mu_c+\frac{z\sqrt{\varepsilon}}{\kappa}.
\]
Each of the 27 settings has 30,000 independent trajectories. Table~\ref{tab:grid} gives the masses and probabilities rounded for display. The CSV files retain the stored precision and integer counts. All 810,000 trajectories reached a stopping threshold before the time limit.

\begin{table}[H]
\centering
\small
\caption{Solvent masses and product probabilities for the three noise strengths; entries are rounded for display. $P_{\rm MC}$ is the measured product probability and $P_{\rm pred}=\Phi(-z)$.}
\label{tab:grid}
\begin{tabular}{rrrrrr}
\toprule
$\varepsilon$ & $z$ & $\mu_2$ & $N$ & $P_{\rm MC}$ & $P_{\rm pred}$\\
\midrule
$10^{-5}$ & -2.0 & 9.930184 & 30000 & 0.97593 & 0.97725 \\
$10^{-5}$ & -1.5 & 9.941617 & 30000 & 0.93450 & 0.93319 \\
$10^{-5}$ & -1.0 & 9.953051 & 30000 & 0.84190 & 0.84134 \\
$10^{-5}$ & -0.5 & 9.964484 & 30000 & 0.69620 & 0.69146 \\
$10^{-5}$ & 0.0 & 9.975918 & 30000 & 0.49963 & 0.50000 \\
$10^{-5}$ & 0.5 & 9.987351 & 30000 & 0.30697 & 0.30854 \\
$10^{-5}$ & 1.0 & 9.998784 & 30000 & 0.15770 & 0.15866 \\
$10^{-5}$ & 1.5 & 10.010218 & 30000 & 0.06357 & 0.06681 \\
$10^{-5}$ & 2.0 & 10.021651 & 30000 & 0.02243 & 0.02275 \\
$2.5\times10^{-6}$ & -2.0 & 9.953051 & 30000 & 0.97877 & 0.97725 \\
$2.5\times10^{-6}$ & -1.5 & 9.958768 & 30000 & 0.93530 & 0.93319 \\
$2.5\times10^{-6}$ & -1.0 & 9.964484 & 30000 & 0.84073 & 0.84134 \\
$2.5\times10^{-6}$ & -0.5 & 9.970201 & 30000 & 0.69150 & 0.69146 \\
$2.5\times10^{-6}$ & 0.0 & 9.975918 & 30000 & 0.49970 & 0.50000 \\
$2.5\times10^{-6}$ & 0.5 & 9.981634 & 30000 & 0.31053 & 0.30854 \\
$2.5\times10^{-6}$ & 1.0 & 9.987351 & 30000 & 0.15900 & 0.15866 \\
$2.5\times10^{-6}$ & 1.5 & 9.993068 & 30000 & 0.06557 & 0.06681 \\
$2.5\times10^{-6}$ & 2.0 & 9.998784 & 30000 & 0.02250 & 0.02275 \\
$10^{-6}$ & -2.0 & 9.961455 & 30000 & 0.97747 & 0.97725 \\
$10^{-6}$ & -1.5 & 9.965071 & 30000 & 0.93097 & 0.93319 \\
$10^{-6}$ & -1.0 & 9.968686 & 30000 & 0.83770 & 0.84134 \\
$10^{-6}$ & -0.5 & 9.972302 & 30000 & 0.68993 & 0.69146 \\
$10^{-6}$ & 0.0 & 9.975918 & 30000 & 0.50110 & 0.50000 \\
$10^{-6}$ & 0.5 & 9.979533 & 30000 & 0.30627 & 0.30854 \\
$10^{-6}$ & 1.0 & 9.983149 & 30000 & 0.15840 & 0.15866 \\
$10^{-6}$ & 1.5 & 9.986764 & 30000 & 0.06723 & 0.06681 \\
$10^{-6}$ & 2.0 & 9.990380 & 30000 & 0.02233 & 0.02275 \\
\bottomrule
\end{tabular}
\end{table}

For setting $i$, let $k_i$ be the product count, $n_i$ the number of trajectories, and $\widehat P_i=k_i/n_i$ the estimated probability. The error bars and the reported goodness-of-fit statistic use binomial variances evaluated at the measured probabilities:
\begin{equation}
\sigma_i^2=\frac{\widehat P_i(1-\widehat P_i)}{n_i},\qquad
\chi^2=\sum_i\frac{[\widehat P_i-\Phi(-z_i)]^2}{\sigma_i^2}.
\label{eq:si-chi}
\end{equation}
We use the full-precision CSV values in these calculations, not the rounded entries in Table~\ref{tab:grid}.

For the 27 probabilities the root-mean-square difference from $\Phi(-z)$ is $1.7281\times10^{-3}$ and the maximum absolute difference is $4.7375\times10^{-3}$. Equation~\eqref{eq:si-chi} gives $\chi^2=26.19499$. With no parameters fitted in the prediction, the large-sample chi-square reference distribution has 27 degrees of freedom and gives $p=0.5078$. The residuals are consistent with the reported sampling uncertainties. The separate contributions are
\[
\begin{array}{c|ccc}
\varepsilon & 10^{-5} & 2.5\times10^{-6} & 10^{-6}\\
\hline
\chi^2 & 12.304 & 7.041 & 6.850
\end{array}
\]
with nine probability estimates in each group. These checks do not show a statistically resolved discrepancy at the tested noise strengths; they do not establish exact equality to the asymptotic probability law.

\section{Independent recovery of the center and width}
\label{sec:si-fit}

The comparison above uses the deterministic predictions without adjustment. We also ask what center and scale are obtained when the stochastic data are fitted to a Gaussian probability curve. For product counts $k_i$ out of $n_i$ trajectories, the binomial log likelihood, apart from constants independent of the fit parameters, is
\[
\ell=\sum_i\left[k_i\log P_i+(n_i-k_i)\log(1-P_i)\right].
\]
The first fit uses $P_i=\Phi[-(\alpha z_i+\beta)]$. The deterministic prediction is $\alpha=1$, $\beta=0$. Maximizing the likelihood gives
\[
\alpha=1.0021662\pm0.0018042,\qquad
\beta=0.0012121\pm0.0017894.
\]
The uncertainties are one standard error, obtained from the inverse expected Fisher information matrix. With $\eta_i=\alpha z_i+\beta$ and $\phi$ the density of a Gaussian with mean zero and variance one, this matrix is
\[
\mathcal I=\sum_i\frac{n_i\phi(\eta_i)^2}{P_i(1-P_i)}
\begin{pmatrix}z_i^2&z_i\\z_i&1\end{pmatrix}.
\]
The coefficient ratio in the notation of Ref.~\cite{Wiggins2026} is $C_{\rm stoch}/C=\alpha^2=1.00434\pm0.00362$, to first order in the fitted uncertainty. This fit is a check of the prediction, not the source of the coefficient used in the Letter.

A second fit permits a common shift $\Delta\mu_c$ in the physical mass threshold:
\[
P_i=\Phi\!\left[-\alpha\kappa\frac{\mu_{2,i}-(\mu_c+\Delta\mu_c)}{\sqrt{\varepsilon_i}}\right].
\]
It gives
\[
\alpha=1.002167\pm0.001804,\qquad
\Delta\mu_c=(-1.73\pm1.83)\times10^{-5}.
\]
A common dimensionless offset $\beta$ and a common mass shift are different models when several noise strengths are fitted together. Both give results consistent with the independently calculated threshold and scale. The verification script and both covariance matrices are included in the Zenodo archive.

\section{Fixed-mass test at small product probability}
\label{sec:si-tail}

\begin{figure}[H]
\centering
\includegraphics[width=0.68\textwidth]{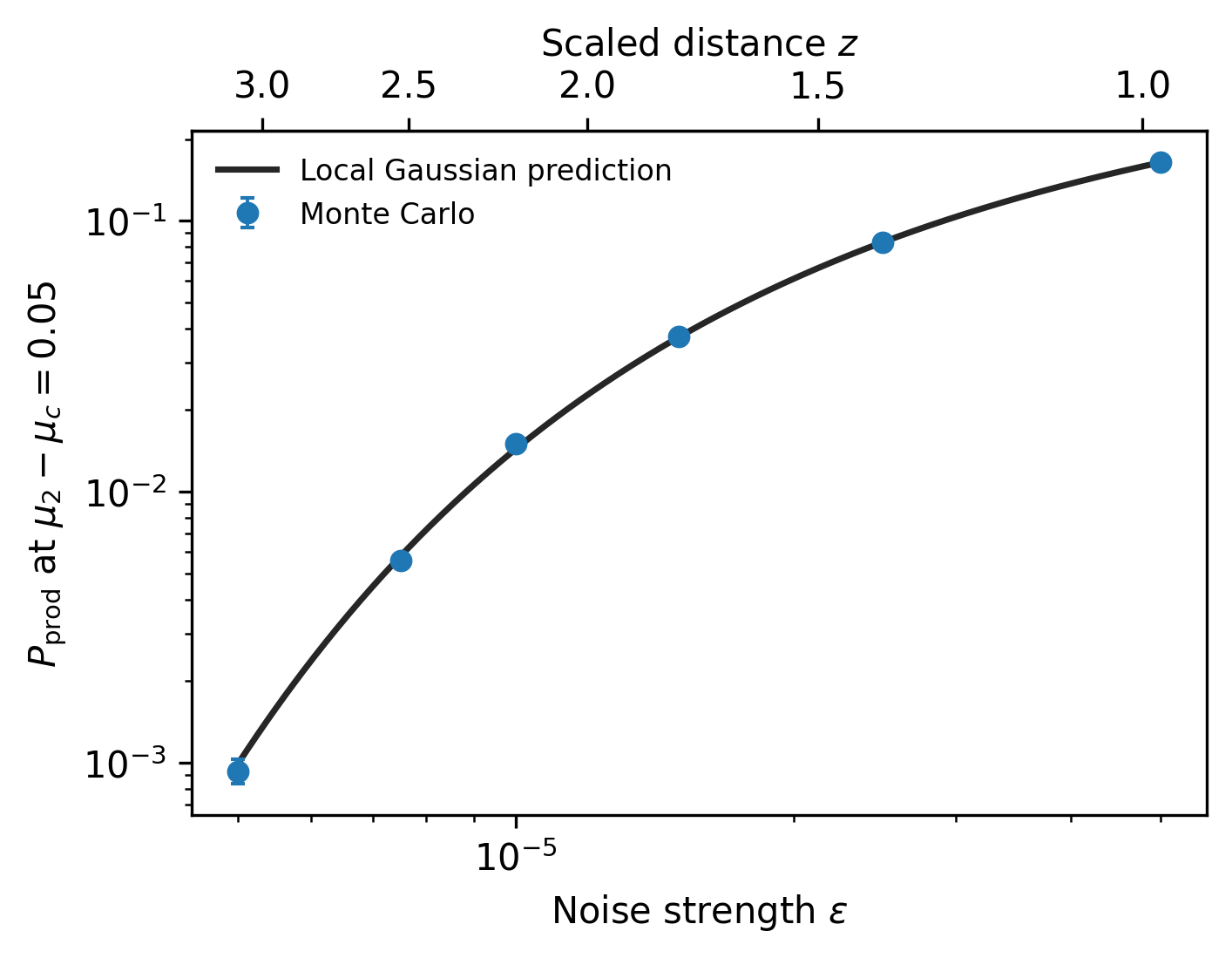}
\caption{Product probability at $\mu_2-\mu_c=0.05$. The mass is fixed while the noise strength is varied. The line is the prediction $\Phi(-z)$; each point uses 100,000 trajectories. Error bars are binomial standard errors evaluated at the measured probabilities. The upper axis gives the scaled distance from the deterministic threshold.}
\label{fig:tail}
\end{figure}

The mass is now held at $\mu_2-\mu_c=0.05$, on the deterministic reactant side, while the noise strength is decreased from $5\times10^{-5}$ to $5\times10^{-6}$. Thus the test varies the probability by changing the forcing rather than by changing the mass. Each of the six settings has 100,000 trajectories. Table~\ref{tab:tail} and Fig.~\ref{fig:tail} give the results. The scaled coordinate ranges from $z=0.978$ to $3.092$.

\begin{table}[H]
\centering
\caption{Product probability at fixed $\mu_2-\mu_c=0.05$.}
\label{tab:tail}
\begin{tabular}{cccc}
\toprule
$\varepsilon$ & $z$ & Monte Carlo & $\Phi(-z)$\\
\midrule
$5.0\times10^{-5}$ & 0.978 & 0.16387 & 0.16407\\
$2.5\times10^{-5}$ & 1.383 & 0.08351 & 0.08335\\
$1.5\times10^{-5}$ & 1.785 & 0.03747 & 0.03710\\
$1.0\times10^{-5}$ & 2.187 & 0.01508 & 0.01439\\
$7.5\times10^{-6}$ & 2.525 & 0.00557 & 0.00579\\
$5.0\times10^{-6}$ & 3.092 & 0.00093 & 0.000993\\
\bottomrule
\end{tabular}
\end{table}

Using the variance convention in Eq.~\eqref{eq:si-chi}, the six probabilities give $\chi^2=4.9559$ and $p=0.5495$, with no fitted parameters. At the smallest noise strength there are 93 product events in 100,000 trajectories: the measured probability is $9.3\times10^{-4}$ and the prediction is $9.93\times10^{-4}$. This extends the numerical test below the probabilities sampled in the central mass-dependent curves.

\section{Time-step convergence}
\label{sec:si-dt}

At $\varepsilon=10^{-6}$, 50,000 trajectories were run at each of four values of $z$ for three time steps. Table~\ref{tab:dt} compares the estimates. No change in probability is resolved beyond the sampling uncertainty at this precision. The test concerns the probability observable, not trajectorywise agreement between integrators.

\begin{table}[H]
\centering
\caption{Time-step convergence of $P_{\rm prod}$ at $\varepsilon=10^{-6}$.}
\label{tab:dt}
\begin{tabular}{ccccc}
\toprule
$z$ & prediction & $\Delta t=0.002$ & $0.001$ & $0.0005$\\
\midrule
$-1$ & 0.84135 & 0.83936 & 0.84358 & 0.84242\\
$0$  & 0.50000 & 0.50066 & 0.50228 & 0.49714\\
$1$  & 0.15866 & 0.15840 & 0.15998 & 0.15892\\
$2$  & 0.02275 & 0.02196 & 0.02246 & 0.02200\\
\bottomrule
\end{tabular}
\end{table}

Across the 45 settings, all 2.01 million trajectories reached a stopping threshold before the time limit in the instrumented replay (Sec.~\ref{sec:si-methods}).

\section{Mean energy input and the weak coupling Langevin limit}
\label{sec:si-energy}

For the additive momentum noise in Eq.~\eqref{eq:sde}, It\^o's formula gives
\[
dH=\sqrt{\varepsilon}\,\frac{p_2}{\mu_2}\,dW_t+\frac{\varepsilon}{2\mu_2}\,dt.
\]
The Hamiltonian terms cancel in this energy balance. The mean input rate is therefore $\varepsilon/(2\mu_2)$. Stopping at a physical threshold or at time $7$ bounds the expected accumulated input by $7\varepsilon/(2\mu_2)$.

The largest such bound among the reported settings occurs for $\varepsilon=5\times10^{-5}$ at $\mu_2=\mu_c+0.05\simeq10.0259$. It is
\[
\Delta E_{\rm mean}\leq1.74548\times10^{-5}<1.8\times10^{-5}.
\]
Relative to $E-V^\ddagger=0.21905267$, the bound is $7.9683\times10^{-5}$, less than $8.0\times10^{-5}$. Most trajectories stop before time $7$, reducing the mean input further. This is a bound on the mean, not on the energy change of an individual realization. In particular, it does not remove energy fluctuations from the dynamics.

For a thermal bath with friction coefficient $\gamma$, temperature $T$, and Boltzmann constant $k_B$, the solvent momentum equation is
\[
dp_2=\left[-\partial_{r_2}V-\gamma p_2/\mu_2\right]dt+\sqrt{2\gamma k_BT}\,dW_t.
\]
Identifying $\varepsilon=2\gamma k_BT$ at fixed positive temperature gives $\gamma=O(\varepsilon)$ as the coupling is reduced. Over a bounded transit time the frictional correction is then $O(\varepsilon)$, smaller than the $O(\sqrt{\varepsilon})$ stochastic displacement used in the probability calculation. In this limit the Hamiltonian threshold and sensitivity give the leading prediction. This is a finite-time weak coupling argument, not a simulation of an equilibrium bath or a result at fixed, non-small friction.

\section{Reproducibility and provenance}
\label{sec:si-repro}

The versioned archive is \url{https://doi.org/10.5281/zenodo.22852306} \cite{Zenodo2026}. It contains the aggregate counts and probabilities, a 45-setting simulation design with seeds, the original numerical kernel and deterministic sensitivity script, portable runners, figure code, and verification records. Individual trajectory time series are not stored; the seeds and numerical code allow the trajectories to be regenerated.

The root \path{README.md} gives installation and execution instructions, and \path{docs/DATA_DICTIONARY.md} defines the CSV fields. The scripts \path{verify_statistics.py}, \path{make_figures.py}, \path{reproduce_simulations.py}, and \path{recompute_coefficient.py}, in the \path{scripts/} directory, reconstruct the statistical results, figures, stochastic experiments, and deterministic coefficient, respectively. New output is written to a separate directory so that the archived data are not overwritten.

The release verification reproduced every archived product count on the documented test platform. A replay with the same method and seeds checks reproducibility and stopping classifications; it is not an independent-integrator validation. The time-step results in Sec.~\ref{sec:si-dt} provide the reported discretization check. Numerical library versions, checksums, and the detailed verification record are included in the archive. Exact binary agreement on other platforms is not guaranteed.

Claude (Anthropic) and ChatGPT (OpenAI) assisted with checking statistical formulas, numerical bounds, and verification and figure code. The reported trajectories were generated by the archived numerical scripts. The author reviewed these checks and verified the reported quantities against the data and code, as stated in the Letter's AI declaration.

\begin{thebibliography}{99}
\small

\bibitem{GarciaCarpenter2019}
R. Garcia-Meseguer, B.K. Carpenter,
Re-Evaluating the Transition State for Reactions in Solution,
Eur. J. Org. Chem. 2019 (2019) 254--266.
\url{https://doi.org/10.1002/ejoc.201800841}.

\bibitem{GarciaCarpenterWiggins2019}
R. Garcia-Meseguer, B.K. Carpenter, S. Wiggins,
The influence of the solvent's mass on the location of the dividing surface for a model Hamiltonian,
Chem. Phys. Lett. 737S (2019) 100030.
\url{https://doi.org/10.1016/j.cpletx.2019.100030}.

\bibitem{PechukasMcLafferty1973}
P. Pechukas, F.J. McLafferty,
On transition-state theory and the classical mechanics of collinear collisions,
J. Chem. Phys. 58 (1973) 1622--1625.
\url{https://doi.org/10.1063/1.1679404}.

\bibitem{PollakPechukas1978}
E. Pollak, P. Pechukas,
Transition states, trapped trajectories, and classical bound states embedded in the continuum,
J. Chem. Phys. 69 (1978) 1218--1226.
\url{https://doi.org/10.1063/1.436658}.

\bibitem{PechukasPollak1979}
P. Pechukas, E. Pollak,
Classical transition state theory is exact if the transition state is unique,
J. Chem. Phys. 71 (1979) 2062--2068.
\url{https://doi.org/10.1063/1.438575}.

\bibitem{Uzer2002}
T. Uzer, C. Jaff\'e, J. Palaci\'an, P. Yanguas, S. Wiggins,
The geometry of reaction dynamics,
Nonlinearity 15 (2002) 957--992.
\url{https://doi.org/10.1088/0951-7715/15/4/301}.

\bibitem{Waalkens2008}
H. Waalkens, R. Schubert, S. Wiggins,
Wigner's dynamical transition state theory in phase space: classical and quantum,
Nonlinearity 21 (2008) R1--R118.
\url{https://doi.org/10.1088/0951-7715/21/1/R01}.

\bibitem{Kramers1940}
H.A. Kramers,
Brownian motion in a field of force and the diffusion model of chemical reactions,
Physica 7 (1940) 284--304.
\url{https://doi.org/10.1016/S0031-8914(40)90098-2}.

\bibitem{GroteHynes1980}
R.F. Grote, J.T. Hynes,
The stable states picture of chemical reactions. II. Rate constants for condensed and gas phase reaction models,
J. Chem. Phys. 73 (1980) 2715--2732.
\url{https://doi.org/10.1063/1.440485}.

\bibitem{Wiggins2026}
S. Wiggins,
Rare fluctuations from normally hyperbolic invariant manifolds,
arXiv:2608.19814 [physics.chem-ph] (2026).
\url{https://arxiv.org/abs/2608.19814}.

\bibitem{Zenodo2026}
S. Wiggins,
Data and code for Solvent Inertia Changes Product Probability by Relocating the Phase-Space Reactivity Boundary,
Zenodo, version 1.0.0 (2026).
\url{https://doi.org/10.5281/zenodo.22852306}.

\bibitem{BartschPRL2005}
T. Bartsch, R. Hernandez, T. Uzer,
Transition state in a noisy environment,
Phys. Rev. Lett. 95 (2005) 058301.
\url{https://doi.org/10.1103/PhysRevLett.95.058301}.

\bibitem{BartschJCP2005}
T. Bartsch, T. Uzer, R. Hernandez,
Stochastic transition states: reaction geometry amidst noise,
J. Chem. Phys. 123 (2005) 204102.
\url{https://doi.org/10.1063/1.2109827}.
\end{thebibliography}
\end{document}